\documentclass[twocolumn,prd,aps,nofootinbib,epsf,floats,amsfonts,amssymb,amsmath]{revtex4}

\usepackage{epsfig}

\usepackage{amssymb}

\newcommand{\be}{\begin{equation}}
\newcommand{\ee}{\end{equation}}
\newcommand{\bea}{\begin{eqnarray}}
\newcommand{\eea}{\end{eqnarray}}
\newcommand{\beaa}{\begin{eqnarray}}
\newcommand{\eeaa}{\end{eqnarray}}
\newcommand{\ba}{\begin{array}}
\newcommand{\ea}{\end{array}}
\newcommand{\bit}{\begin{itemize}}
\newcommand{\eit}{\end{itemize}}
\newcommand{\ben}{\begin{enumerate}}
\newcommand{\een}{\end{enumerate}}

\def\lab{\label}
\def\lan{\langle}

\def\lf{\left}

\def\pa{\partial}
\def\ran{\rangle}
\def\rar{\rightarrow}

\def\ri{\right}

\def\Ga{\Gamma}

\def\la{\lambda}

\begin{document}


\title{Vortices in brain waves}

\author{Walter J. Freeman${}^{a}$ and Giuseppe Vitiello${}^{b}$
\vspace{3mm}}

\address{${}^{a}$ Department of Molecular and Cell Biology\\
University of California, Berkeley CA 94720-3206
USA\\
dfreeman@berkeley.edu - http://sulcus.berkeley.edu
\\ 
${}^{b}$ Dipartimento di Matematica e Informatica and INFN\\
Universit\`a di Salerno,
I-84100 Salerno, Italy\\
vitiello@sa.infn.it - http://www.sa.infn.it/giuseppe.vitiello}


\vspace{18mm}



\begin{abstract}
Interactions by mutual excitation in neural populations in human and
animal brains create a mesoscopic order parameter that is recorded
in brain waves (electroencephalogram, EEG). Spatially and spectrally
distributed oscillations are imposed on the background activity by
inhibitory feedback in the gamma range (30-80 Hz). Beats recur at
theta rates (3-7 Hz), at which the order parameter transiently
approaches zero and microscopic activity becomes disordered. After
these null spikes, the order parameter resurges and initiates a
frame bearing a mesoscopic spatial pattern of gamma amplitude
modulation that governs the microscopic activity, and that is
correlated with behavior. The brain waves also reveal a spatial
pattern of phase modulation in the form of a cone. Using the
formalism of the dissipative many-body  model of brain, we describe
the null spikes and the accompanying phase cones as vortices.
\end{abstract}


\maketitle

\section{Introduction}

The dissipative quantum model of brain predicts two main features of
neurophysiological data \cite{11}: the coexistence of physically
distinct amplitude modulated (AM) and phase modulated (PM) patterns
correlated with categories of conditioned stimuli and the remarkably
rapid onset of AM patterns into irreversible sequences that resemble
cinematographic frames. These features of the brain activity are
observed in laboratory by means of imaging of scalp potentials
(electroencephalograms, EEGs) and of cortical surface potentials
(electrocorticograms, ECoGs) of animal and human from high-density
electrode arrays. The mesoscopic neural activity of neocortex
appears indeed consisting of the dynamical formation of spatially
extended neuronal domains in which widespread cooperation supports
brief epochs of patterned synchronized oscillations, which have been
demonstrated to occur in the $12-80 ~Hz$ range ($\beta$ and $\gamma$
ranges). They re-synchronize in frames at frame rates in the $3-12
~Hz$ range ($\theta$ and $\alpha$ ranges) \cite{7,8,9,10,11}. These
patterns, or ``packets of waves'', appear often to extend over
spatial domains covering much of the hemisphere in rabbits and cats
\cite{12,13}, and over the length of a $64 \times 1$ linear $19 ~cm$
array \cite{7} in human cortex with near zero phase dispersion
\cite{14,15}. Synchronized oscillation of large-scale neuronal
assemblies in $\beta$ and $\gamma$ ranges have been detected also by
magnetoencephalographic (MEG) imaging in the resting state and in
motor task-related states of the human brain \cite{Bassett}. The
patterns of phase-locked oscillations are intermittently present in
resting, awake subjects as well as in the same subject actively
engaged in cognitive tasks requiring interaction with environment,
so they are best described as properties of the background activity
of brains that is modulated upon engagement with the surround.

Neither the electric field of the extracellular dendritic current
nor the extracellular magnetic field from the high-density electric
current inside the dendritic shafts, which are much too weak, nor
the chemical diffusion, which is much too slow, appear to be able to
fully account for the observed cortical collective activity
\cite{11,FreemanNDN}. On the contrary, it turns out that the
many-body dissipative model \cite{Vitiello:1995wv} is able to
account for the dynamical formation of synchronized neuronal
oscillations \cite{11}.  This will not be illustrated again here
(the reader may find detailed discussion in
\cite{11,Vitiello:1995wv,11a}). We only recall that each AM pattern
is described to be consequent to spontaneous breakdown of symmetry
triggered by external stimulus \cite{UR,Vitiello:1995wv} and is
associated with one of the quantum field theory (QFT) unitarily
inequivalent ground states \cite{11,Vitiello:1995wv}. Their
sequencing is associated to the non-unitary time evolution implied
by dissipation \cite{Vitiello:1995wv,11}. In this paper we focus our
attention on a crucial neural mechanism, that we deduced from
experimental observations of a pattern called ``Coordinated Analytic
Phase Differences" (CAPD) \cite{8,9,10,12,13}, consisting in the
fact that the event that initiates the transition to a perceptual
state is an abrupt decrease in the analytic power of the background
activity to near zero, depicted as a null spike, associated with the
concomitant increase of spatial variance of analytic phase. The null
spikes tend to recur aperiodically at rates in the theta  ($3-7
~Hz$) and alpha ($8-12 ~Hz$) ranges. By use of the Hilbert
transform, the local structure of CAPD is visualized in the real and
imaginary parts, $a(x)$ and $b(x)$, respectively, of the ECoG
sampled wave function $\psi (x)$ in the selected spectral pass band
\be \lab{psi} \psi (x) = {\bf A}^{2} (x) e^{i \phi (x)} ~, \ee
where $x \equiv (x,y,t)$ in the two surface dimensions of cortex
($3$ dimensions for the microscopic level of networks), and the
analytic power ${\bf A}^{2}(x)$ and  the analytic phase $\phi (x)$
are
\bea  {\bf A}^{2}(x) &=& \sqrt{a^{2}(x) + b^{2}(x)} ~,\\
 \phi (x) &=& \arctan \frac{b(x)}{a(x)} ~, \eea
respectively.  During periods of high amplitude the spatial
deviation of phase ($SD_{X}$) is low and the phase spatial mean
tends to be constant within frames and to change suddenly between
frames, indicating coherence and coordinated phase differences.
${\bf A}^{2}(x)$ forms a feature vector that serves as our order
parameter (see below and Refs. \cite{11,11a}).

The reduction in the amplitude of the spontaneous background
activity induces a brief state of indeterminacy in which the
significant pass band of the electrocorticogram (ECoG) is near to
zero and the phase of ECoG is undefined (Figure 1).

\vspace{4mm}



\centerline{\epsfysize=2.7truein\epsfbox{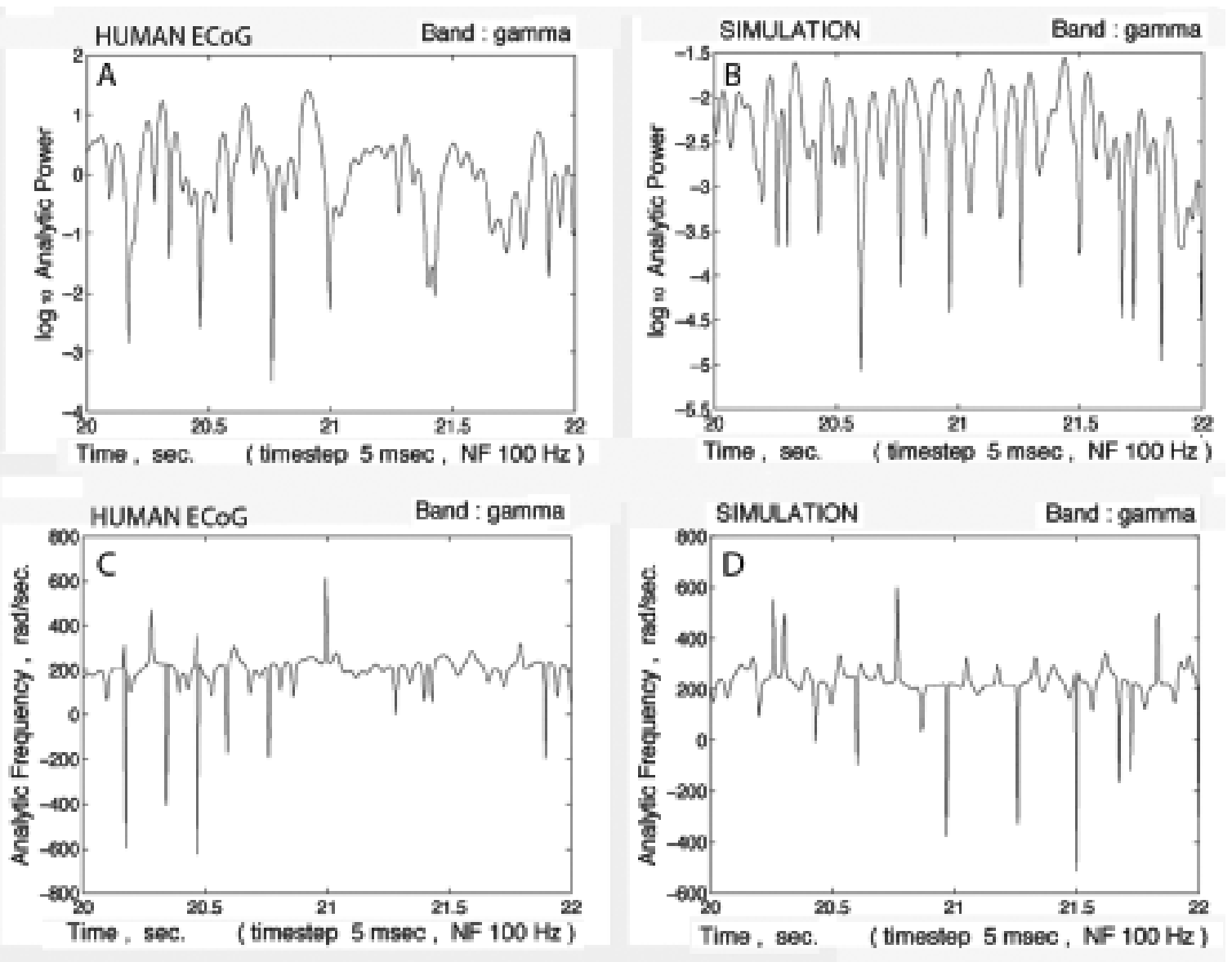}}
\vspace{.3cm}
{\small \noindent Figure 1. The temporal patterns of null spikes are
illustrated; each spike initiates a spatial phase cone. A. The
logarithm of the analytic power (four ECoG signals superimposed from
an 8x8 array) in the gamma range (25-50 Hz) shows the downward null
spikes demarcating onsets of cones at irregular intervals. C. The
spikes in analytic phase coincide with the null spikes in power; the
differences between signals reflect the high spatial variance
contributed by the cones. B. The statistical properties of null
spikes are replicated by cumulatively summing Gaussian noise and
applying to the signal the same band pass filter (1/4 to 1/2 the
Nyquist frequency, 100 Hz). D. The spikes in analytic phase coincide
with the null spikes in power.
 }

\vspace{4mm}

The cortex can be driven across such a ``phase transition" process
to a new AM pattern by the stimulus arriving at or just before this
state. The observed velocity of spread of phase transition is
finite, i.e. there is no ``instantaneous" phase transition.
Experimental evidence of CAPD over large cortical areas indicates
that the neuronal correlation length would cover an entire cerebral
hemisphere virtually instantaneously (practically without delay in
the gamma activity), if measured at the critical transition. Between
the null spikes the cortical dynamics is (nearly) stationary for
$\sim 60-160 ~ms$. This is called a frame. The transitions by which
they form are shorter by an order of magnitude.

In this paper we discuss such a mechanism  and describe, in the
formalism of the dissipative model, the observed occurrence of phase
cones and the dynamic formation of vortices in brain waves. The
phase cone is a spatial phase gradient that is imposed on the
carrier wave of the wave packet in a  frame by the propagation
velocity of the largest axons having the highest velocity in a
distribution. The  location of the apex is a random variable across
frames that is determined by the accidents of where the null spike
is lowest and the background input is highest. The null spike has
rotational energy at the geometric mean frequency of the pass band,
so it is called a vortex. The vortex occupies the whole area of the
phase-locked neural activity of the cortex for a point in time. One
more observed feature is the random variation from each frame to the
next of the slope of the conic phase gradient, negative with
explosion, positive with implosion. The negative gradient could be
explained in conventional neurodynamics (e.g. in terms of a
pacemaker), but not the positive gradient. Also, there is no
explanation in the conventional framework of why both gradients, the
positive and the negative one, occur. These features have been
documented as markers of the interface between microscopic and
mesoscopic phenomena.

In the process of non-instantaneous phase transitions (as those
observed in brain) the dissipative model predicts the existence of
vortex singularity associated (at the vortex core) with the abrupt
decrease (null spike) of the order parameter (the analytic
amplitude) and the concomitant increase of spatial variance of the
phase field (the analytic phase). The resulting phase cones present
both phase gradients, the positive and the negative one, as in the
observations.

Phase transitions and vortex solutions in the dissipative model are
discussed in Section II where it is shown how the model predicts the
observed feature of null spikes. Heat dissipation involved in the
disappearance/emergence of coherence is discussed in Section III. In
Section IV it is presented the discussion of size, number and time
dependence of transient non-homogeneous structures appearing during
non-instantaneous phase transitions, such as those observed in
brain. The formation of imploding and exploding phase cones is shown
to be allowed, as indeed deduced from observations. Section V is
devoted to final remarks and conclusions.

\section{Phase transitions, vortex solutions and null spikes}

We start by recalling that in the dissipative model of brain
spontaneous breakdown of the rotational  symmetry of electrical
dipoles of water and other molecules \cite{Jibu,Vitiello:1995wv}
implies the existence of Nambu-Goldstone modes (NG)
\cite{Goldstone:1961eq,ITZ} which in such a context have been called
the dipole wave quanta (DWQ), say $P(x)$ and $P^{\dag}(x)$. The
non-vanishing polarization density ${\cal P} = \rho \delta$, where
$\rho$ and $\delta$ are the charge density and the (average) dipole
length, is expressed in terms of these field modes
\cite{DelGiudice:1985} and the system ground state is obtained in
terms of coherent condensation of the DWQ. One then considers the
spontaneous breakdown of the phase
symmetry 
and the charge density wave function $\sigma (x)$ is written
\cite{DelGiudice:1985} as
\be\lab{vs9a} \sigma (x) = \sqrt{\rho (x)} e^{i\theta(x)}  ~, \ee
with real $\rho (x)$ and $\theta(x)$.  The ``phase" $\theta(x)$ is
the NG field associated with the breakdown of global phase symmetry.
The boson condensation of the field $\theta(x)$ in the system ground
state is formally described by the transformation
\be\lab{vs9} \theta(x) \rar \theta(x) - \frac{e_0 {v}^2}{Z} f(x) ~.
\ee
The c-number condensation function $f(x)$ satisfies the same
equation satisfied by the $ \theta(x)$ field, i.e. $\pa^2 f(x) =0$.
The constant $Z$ is the wave function renormalization constant,
$e_0$ and ${v}$ are the electron charge and the constant entering
the symmetry breakdown condition $<0| \rho (x)|0> = v \neq 0$
\cite{DelGiudice:1985,Matsumoto:1975fi,Umezawa:1993yq}.

Coherent domains of finite size are obtained by non-homogeneous
boson condensation. The condensate is described by the function
$f(x)$ which acts as a ``form factor" specific for the considered
domain
\cite{Umezawa:1993yq,Alfinito:2000ck,Alfinito:2002a,Alfinito:2002b}.
One can show \cite{DelGiudice:1985,Matsumoto:1975fi} that
mathematical consistency requires that the electromagnetic vector
potential $a_\mu(x)$ has then to satisfy the equation
\be\lab{g3.vs20} (\pa^2 + m_V^2) a_\mu(x) \, = \,\frac{m_V^2}{ e_0}
\pa_\mu f(x)~. \ee
We adopt the gauge condition $\pa^{\mu} a_\mu(x) = 0$. Eq.
(\ref{g3.vs20}) is the classical Maxwell equation for the massive
vector potential $a_{\mu}$ ($m_V$ is its mass). The classical ground
state current $j_{\mu,cl}$ turns out to be
\be\lab{g3.vs21} j_{\mu,cl}(x)\equiv \lan 0| j_{\mu}(x) |0 \ran \,
=\,
 m_V^2 \lf[ a_\mu(x) - \frac{1}{e_0} \pa_\mu f(x) \ri]~,
\ee
and we have $\pa^{\mu}j_{\mu,cl}(x) = 0$. The term $ m_V^2 a_\mu(x)$
is the well known {\em Meissner current}, while $ \frac{m_V^2}{e_0}
\pa_\mu f(x)$ is the {\em boson current}.

{\it The mesoscopic field and current are thus given in terms of the
boson transformation function}. It is also remarkable that the
classical current is related with ${\pa}_{\mu}f$, i.e. with
variations in the boson transformation function.

The important point is that such a condensation function $f(x)$ has
to carry some topological singularity in order for the condensation
process to be physically detectable. The function $f(x)$ carrying a
topological singularity is not single-valued and thus is
path-dependent:
\be \lab{g3.ts1} [\pa_\mu,\pa_\nu]\,f(x) \neq 0~, \quad  for
\;certain \; \mu\, , \, \nu \, , \, x . \ee
On the other hand, observables may be influenced by gradients in the
Bose condensate and thus $\pa_\mu \, f$ is related with observables
and therefore has to be single-valued, i.e.
$[\pa_\rho,\pa_\nu]\,\pa_\mu f(x)\,=\,0$. A regular function $f(x)$
would produce a condensation which could be easily ``gauged" away by
a convenient field transformation. From (\ref{g3.vs20}) we obtain
$a_{\mu}(x) = \frac{1}{\pa^2 + m_V^2}\pa_\mu \, f(x)$. When $f(x)$
is regular, this gives $\pa^2 a_{\mu}(x) = 0$ since $\pa^2 f(x) =
0$. Thus Eq. (\ref{g3.vs20}) implies $a_{\mu}(x) = \frac{1}{e_{0}}
\pa_{\mu} f(x)$ for regular $f(x)$, which in turn implies zero
classical field ($F_{\mu\nu} = \pa_{\mu} a_{\nu} - \pa_{\nu}
a_{\mu}$) and zero classical current ($j_{\mu,cl} = 0$) since the
Meissner and the boson current cancel each other.
It is indeed well known \cite{Anderson:1984a} that the gauge field
$a_{\mu}$ is expelled out of the ordered domain region (it is there
vanishing) where the order parameter is of course non-zero and
$f(x)$ does not have singularities. On the contrary, the gauge field
is non zero in the regions where $f(x)$ presents non-trivial
topological singularities such as line singularities, e.g. on the
line $r = 0$ in the core of a vortex: we have there the "normal"
(disordered) state rather than the ordered one and the non-vanishing
massive gauge field there propagates (the Anderson-Higgs-Kibble
mechanism) \cite{Anderson:1984a,hig}. On the boundaries between the
normal and the ordered regions the phase field gradients are
non-zero. Instead they are zero in the normal region, i.e. in the
vortex core. Consistently with this scenario, one can also show
\cite{Umezawa:1993yq,Alfinito:2002a,Alfinito:2002b} that the phase
transition from one state space to another (unitarily inequivalent)
one can be only induced by a singular boson transformation function
$f(x)$. This is the reason why topologically non-trivial extended
objects, such as vortices, appear in the processes of phase
transitions \cite{Umezawa:1993yq,Alfinito:2002a,Alfinito:2002b}.
Stated in different words, this means that phase transitions driven
by boson transformations are always associated with some
singularities in the field phase. We thus recognize that in the
brain the null spike (the observed abrupt decrease in the order
parameter and the concomitant increase in the phase field gradients
in the phase transition from an AM amplitude to another one) is
indeed characterized by the topological singularity of the function
$f(x)$. In the case of phase symmetry summarized above, the
stationary function $f(x)$ solution of our problem carries a vortex
singularity and is given by
\be \lab{g3.ts1vo} f(x) = \arctan \left(\frac{x_{2}}{x_{1}}\right)~.
\ee
Eq. (\ref{g3.ts1vo}) shows that the phase is undefined on the line
$r = 0$, with $r^{2} = x^{2}_{1} + x^{2}_{2}$, consistently with the
observed phase indeterminacy in the process of transition between
two AM pattern frames. As usual in these cases, as a result of the
single-valuedness of $\sigma (x)$ the topological singularity is
characterized by the winding number $n$: $\oint \nabla f \cdot dl =
\frac{2\pi}{e_{0}} n$, $n = 0, \pm 1, \pm 2, ...$, when the
integration is performed along the closed circle $(0, 2\pi)$ (flux
quantization).

The stimulus arriving at or just before the abrupt decrease of
analytic power drives the cortex across the phase transition process
to the new AM pattern. As remarked elsewhere \cite{11}, the
dissipative model predicts that the response amplitude depends not
on the input amplitude, but on the intrinsic state of the cortex,
specifically the degree of reduction in the power and order of the
background brown noise. As a matter of fact, such a feature is one
of the merits of the mechanism of spontaneous breakdown of symmetry
where the external stimulus (as in the case of the brain) only acts
as a trigger, the correlated phase regime being reached as the
effect of the system inner dynamics. This explains the observed lack
of invariance of AM patterns with invariant stimuli
\cite{11,9,10,12,13}. The power is indeed not provided by the input,
exactly as the dissipative model predicts, but by the pyramidal
cells.

We also observe that the initial site where non-homogeneous
condensation starts (the phase cone apex) is not conditioned by the
incoming stimulus, but is randomly determined by the concurrence of
a number of local conditions, such as where the null spike is lowest
and the background input is highest, in which the cortex finds
itself at the transition process time. The apex is never initiated
within frames (in the broken symmetry phase or ordered region), but
between frames (during phase transitions), as it is indeed predicted
by the dissipative model (vortices occur during the critical regime
of phase transitions). The null spike appears in the band pass
filtered brown noise activity and can be conceived as a {\it
shutter} that blanks the intrinsic background ECoG. When the order
parameter goes to zero the microscopic activity (of the background
state) does not decrease but, consistently with the model
description, it becomes disordered, unstructured (fully symmetric).
In such a state of very low analytic amplitude, the analytic phase
is undefined, as it is indeed at the center line of the vortex core,
and the system, under the incoming weak sensory input, may re-set
the background activity in a new AM frame, if any, formed by
reorganizing the existing activity, not by the driving of the
cortical activity by input (except for the small energy provided by
the stimulus that is required to force the phase transition). The
analytic amplitude decrease repeats in the theta or alpha range,
independently of the repetitive sampling of the environment by
limbic input. Consistently with observations, in the dissipative
model the reduction in activity constitutes a singularity in the
dynamics at which the phase is undefined. The aperiodic shutter
allows opportunities for phase transitions.

\section{Heat dissipation and disappearence/emergence of coherence}

We have already commented upon the remarkable interplay between the
emergence of mesoscopic field and currents and the microscopic
phenomenon of boson condensation (cf. the discussion after Eqs.
(\ref{g3.vs20}) and (\ref{g3.vs21})). We further observe that the
neural mechanism of perception depends on repeated transfer of
mesoscopic energy to microscopic energy and vice-versa as the basis
for the disintegration of a mesoscopic AM pattern and the formation
of a new one, respectively. In the dissipative model these energy
transfers are controlled by the time derivative of the number, $\dot
N$, of the $\theta$ phase field condensate
\cite{Vitiello:1995wv,11}:
\be d E = \sum_{k} E_{k} \dot{\cal N}_{k} d t =
 {1\over{\beta}} d {\cal S}  ~.
  \lab{(23)}
\ee
Eq. (\ref{(23)}) holds provided changes in the inverse temperature
$\beta$ are slow, which is what actually happens in mammalian brain
which keep their temperature nearly constant. It relates the changes
in the energy $ E \equiv \sum_{k} E_{k} {\cal N}_{k}$ and in the
entropy ${\cal S}$ implied by the minimization of the free energy at
any $t$. Here $E_{k}$ and ${\cal N}_{k}$ denote the energy and the
number of the NG phase field excitations of momentum $k$. As usual
heat is defined as $ {dQ={1\over{\beta}} dS}$. We thus see how,
through the variations in time of the phase field condensate, the
entropy changes and heat dissipation involved in the
disappearance/emergence of the coherence (ordering) associated to
the AM patterns turns into energy changes. Heat dissipation appears
indeed to be a significant variable in laboratory observations. We
remark that, consistently with observations, the variations
$\pa_{\mu} f$ of the $\theta$ phase field condensate is detectable
at the mesoscopic level only by the variations of the analytic
phase.

Also concerning the mesoscopic/microscopic interplay, it has to be
remarked that the vortex solution in the dissipative model, although
is dynamically generated through the non-homogeneous boson
condensation mechanism, which is a truly quantum mechanism,
manifests itself as a solution of  non-linear {\it classical}
equations.
This a general feature of QFT, where many kinds of topologically
non-trivial solutions of classical field equations (soliton
solutions) are described as mesososcopic ``envelops" of microscopic
boson condensates (for a detailed discussion on the
quantum/classical interplay in field theories with topologically
non-trivial solutions see \cite{Jackiw:1977a}; see also
\cite{Umezawa:1993yq,Matsumoto:1975fi,Vitiello:2001}). The
dissipative quantum model of brain thus provides classical
mesoscopic phenomena originated form the underlying quantum
dynamics. As elsewhere stressed \cite{11,11a,Vitiello:1995wv}, in
such a model the neurons, the glia cells and other physiological
units are {\it not} quantum objects. The quantum degrees of freedom
are those associated to the dipole vibrational field and to other
fields such as the phase field.

Summarizing, the spatial gradient of $f(x)$ in the Bose condensate
of the $\theta (x)$ ``phase" field accounts for the phase cone which
is indeed a spatial phase gradient imposed on the carrier wave of
the wave packet. The vortex solution  arises as an effect of
non-homogeneous condensation of the phase field $\theta (x)$, which
spans (almost) the whole system since it is a (quasi-)massless field
(it is a collective mode). This explains the fact that in its
life-time the vortex is observed to occupy the whole area of the
phase-locked neural activity of the cortex. In this connection, it
is interesting to comment on the size, number and time dependence of
transient non-homogeneous structures appearing during
non-instantaneous phase transitions (lasting a finite time
interval), such as those observed in brain. We consider this in the
next Section and we find that converging and diverging (imploding
and exploding) phase cones are formed, as indeed deduced from
observations.

\section{Phase cones and critical regime in the dissipative model}

Transition processes occurring in a finite span of time in which the
formation of defects (e.g. vortex strings) occurs, have been studied
by numerical simulations and theoretical modeling in a number of
problems of physical interest
\cite{Bettencourt,Hindmarsh,Alfinito:2002b}. In these processes, a
maximally stable new configuration is attained after a certain lapse
of time since the transition has started. The system is said to be
in the critical or Ginzburg regime during such a lapse of time. In
the critical regime one deals with the matter  field (the Higgs
field in elementary particle physics) condensate and the NG ($\theta
(x)$ phase field) condensate. We have considered the last one in the
discussion above. Enough reliable information on the critical regime
behavior of the former one is provided by using the harmonic
approximation for the evolution of the order parameter $v$, which is
now assumed to be space-time dependent (non-homogeneous Higgs
condensate) \cite{Bettencourt,Alfinito:2002a,Alfinito:2002b}. By
resorting to such an approximation in our present brain problem, we
expand the $v$ field into partial waves:
\be \lab{g5.17} v(x,t) = \sum_{\bf k} \ \{u_{{\bf k} }(t) e^{i{\bf
k}\cdot {\bf x}}+u_{{\bf k}}^{\dag}(t)e^{-i{\bf k}\cdot{\bf x} }\}
~.
\ee
In general, $v$ and $u$ depend also on the temperature.
However, we will omit the dependence on temperature
since this does not affect our discussion. In the harmonic potential
approximation of the the Ginzburg-Landau (GL) formalism we have the
equations for the parametric oscillators $u_{\bf k}$
\cite{Perelomov:1986tf} (see also
\cite{Alfinito:2002a,Alfinito:2002b}) for each $k$-mode ($k \equiv
\sqrt {{\bf k}^{2}}$):
\be \stackrel{..}{u}_{\bf k}(t)+ ({\bf k}^{2} - m^{2})u_{\bf k}=0
~.\lab{g5.19} \ee
Note that the sign of the mass term $m^{2}$ is consistent with the
occurrence of spontaneous breakdown of symmetry
\cite{Alfinito:2002b,Manka:1990}. The oscillator frequency is
\be M_{k}(t)\ =\ \sqrt { {\bf k}^{2} - m^{2}(t) }. \lab{g5.20} \ee
$M_{k}(t)$ is required to be real for each $k$ and, in full
generality, in Eq. (\ref{g5.20}) we are assuming that $m^{2}$ may
depend on time. The reality condition on $M_{k}(t)$  is satisfied
provided at each $t$, during the critical regime time interval, it
is
\be {\bf k}^{2} \ge m^{2}(t)~, \lab{g5.21} \ee
for each $k$-mode. This turns out to be a condition on the $k$-modes
propagation. Let $t=0$ and $t= \tau$ denote the times at which the
critical regime starts and ends, respectively. For a given ${\bf
k}$, Eq. (\ref{g5.21}) holds up to a time $\tau_{k}$ after which
$m^{2}(t)$, for $t > \tau_{k}$, is larger than ${\bf k}^2$. The
corresponding $k$-mode can propagate in a span of time $0\le\ t \
\le \tau_k$. Thus the ``effective causal horizon" \cite{kib2,zurek1}
can happen to be inside the system (possible formation of more than
a domain) or outside (single domain formation) according to the time
occurring for reaching the boundaries of the system is longer or
shorter than the allowed propagation time. This determines the
dimensions to which the domains can expand.

The value of $\tau_k$ is given when the explicit form of $m^2 (t)$
is assigned. One may then consider to model the time dependence of
$m(t)$ \cite{Alfinito:2002b} in a way to allow defect (i.e. vortex)
formation. We thus choose $m^2 (t)$ to be:
\be  m^{2}(t)  = m_{0}^{2} \ e^{2h(t)} ~. \lab{g5.151} \ee
The function $h(t)$ is assumed to be positive, monotonically growing
in time from $t = 0$ to $t = \tau$. The correlation propagation time
is implicitly given by:
\be h(\tau_k) =  \ln\left(k  \over m_{0} \right) \; \propto  \;
\ln\left( L\over \xi \right)~, \lab{g5.221} \ee
$h(\tau_k)$ resembling the commonly called string tension
\cite{zurek1}.  In Eq.  (\ref{g5.221}), $\xi$ is the correlation
length corresponding to the $k$-mode propagation and $L \propto
m_{0}^{-1}$. $L$ acts as an intrinsic infrared cut-off. Small $k$
values are indeed excluded, due to Eq. (\ref{g5.21}), by the
non-zero minimum value of $m^2$.  Correspondingly, long wave-lengths
are precluded, i.e. only domains of finite size can be obtained. At
the end of the critical regime the correlation may extend over
domains of linear size of the order of $\lambda_{k} \propto
m^{-1}(\tau)$.

Our model is further specified by choosing an explicit analytic
expression for $h(t)$.  We choose \cite{Alfinito:2002b}:
\be h(t)= \pm \frac{a t}{b t^{2} + c}~, \lab{g5.24a} \ee
where $a, b, c$ are (positive) parameters chosen so to guarantee the
correct dimensions and the correct behavior in time. We denote their
ratios by  $c/a\lambda \equiv  \tau_{Q}$, $a\lambda/b \equiv
\tau_{0}$, with $\lambda$ an arbitrary constant. We note that
$h(\tau_{Q}) = h(\tau_{0})$. The time derivative of $h(t)$, and thus
of $m^{2}(t)$, is zero at $t = \tau = \pm \sqrt{\tau_{Q}\tau_{0}}$.
$\tau$ thus plays the role of the equilibrium time scale. We observe
that
\be h(t) = \pm  \frac{1}{\la \tau_{Q}} \frac{1}{1 +
\frac{t^{2}}{\tau^{2}}} ~t \approx \pm \frac{\Ga}{2} ~t~,
\lab{g5.24} \ee
for $t^{2}/\tau^{2} \approx 1$, with $\Ga \equiv 1/\la \tau_{Q}$.

The number of defects (of vortices) $n_{def}$  possibly appearing
during the critical regime is given in the linear approximation by
\cite{Alfinito:2002b,zurek1}:
\be n_{def} \propto m^{2}(\tau) 
\approx m_{0}^{2}\ | \tau/{\lambda \tau_{Q}}|~.
\lab{g5.24ter} \ee
%

We observe that the size of the vortex core is given by
$(m(t))^{-1}$ and thus Eqs. (\ref{g5.151}) and (\ref{g5.24}) show
that such a size evolves in time as $e^{\mp \Ga ~t}$, $t < \tau$ ($t
< \tau_{k}$ for the $k$-mode). This means that we have both,
converging (imploding) and diverging (exploding) regimes, as indeed
found in laboratory observations of the phase cone behaviors. Since
the ``normal" state is confined to the vortex core, the shrinking of
such a region (imploding regime) may signal that long range
correlation, i.e. ordering, is prevailing (the vortex is ``squeezed
out"); in the opposite case of enlargement of the vortex core
(exploding regime), local correlations (disorder) prevail. This
agrees with the postulate reached on the basis of laboratory
observations
according to which implosion or explosion is obtained if the long
axon connections or the local connections predominate, respectively
\cite{38}.

\section{Final remarks and conclusion}

As a final comment we remark that Eq. (\ref{g5.24}) shows that the
$\pm$ signs in Eq. (\ref{g5.24a}) amount to  working with both
elements of the basis $(e^{{+\Ga}/{2} ~t}, e^{{-\Ga}/{2} ~t})$, as
indeed required by mathematical correctness. In this sense,  the
$\pm$ double sign cannot be avoided in the model choice of $h(t)$.
From a physical point of view, it is equivalent to working with time
evolution pointing in one given time direction (say the $t > 0$
arrow of time) and with its ``time-reversed" copy or image. This is
perfectly consistent with one of the main features of the
dissipative model where time-reversed excitations are introduced,
thus ``doubling" the system degrees of freedom \cite{Jackiw2}, so
that one is led to consider the time-reversed image of the system,
its ``{\it Double}". It is interesting to observe that such a model
feature finds a connection with the laboratory observation of the
exploding/imploding feature in the phase cone behavior. The
description of the vortex singularities appearing in the process of
phase transitions turns out to be crucial in the understanding of
the nature of the engagement of the subject with the environment in
the action-perception cycle. By the continual updating of the
meanings of the flows of information exchanged  in its relation with
the environment, the brain proceeds from information to knowledge in
its own world as it is known by itself that we describe as its
Double \cite{Vitiello:2001}.

\section*{Acknowledgments}

The authors thank  Bill Redfearn, Sean O'Nuillain and José Rodriguez
for technical assistance. Partial financial support by INFN and MIUR
is also acknowledged.



\end{document}